\documentclass[prd,preprint,floats,floatfix,showpacs,nofootinbib,superscriptaddress]{revtex4}
\usepackage{amssymb,amsmath,graphicx,amscd,xcolor,amsthm}

\usepackage{graphicx}
\usepackage{dcolumn}
\usepackage{verbatim} 
\usepackage{bm}
\usepackage[utf8]{inputenc} 
\usepackage{graphics}
\usepackage{slashed}
\usepackage{amssymb}
\usepackage{natbib}
\usepackage{amsmath}
\usepackage{amsthm}
\usepackage{url}
\usepackage{color}
\newcommand{\fracc}[2]{\frac{\textstyle{#1}}{\textstyle{#2}}}

\newcommand{\p}{\partial}
\begin{document}

\title{Controlled opacity in a class of nonlinear dielectric media}

\author{E. \surname{Bittencourt}}
\email{bittencourt@unifei.edu.br}
\affiliation{Instituto de F\'{\i}sica e Qu\'{\i}mica,    Universidade Federal de Itajub\'a,\\
Itajub\'a, Minas Gerais 37500-903, Brazil}
\author{G. H. S. \surname{Camargo}}
\email{guilhermehenrique@unifei.edu.br}
\affiliation{Instituto de F\'{\i}sica e Qu\'{\i}mica, Universidade Federal de Itajub\'a,\\
Itajub\'a, Minas Gerais 37500-903, Brazil}
\author{V. A. \surname{De Lorenci}}
\email{delorenci@unifei.edu.br}
\affiliation{Instituto de F\'{\i}sica e Qu\'{\i}mica, Universidade Federal de Itajub\'a,\\
Itajub\'a, Minas Gerais 37500-903, Brazil}
\author{R. \surname{Klippert}}
\email{klippert@unifei.edu.br}
\affiliation{Instituto de Matem\'atica e Computa\c{c}\~ao, Universidade Federal de Itajub\'a,\\
Itajub\'a, Minas Gerais 37500-903, Brazil}


\date{\today}

\begin{abstract}
Motivated by new technologies for designing and tailoring metamaterials, we seek properties for certain classes of nonlinear optical materials that allow room for a reversibly controlled opacity-to-transparency phase transition through the application of external electromagnetic fields. We examine some mathematically simple models for the dielectric parameters of the medium and compute the relevant geometric quantities that describe the speed and polarization of light rays.
\end{abstract}

\maketitle

\section{Introduction}
Driven by the recent advances in designing and tailoring new (meta)materials \cite{meta}, there has been increasing interest in studying new optical phenomena in recent years. In order to remind readers of some remarkable achievements of such investigations, we mention the interesting effects of electromagnetically induced transparency (EIT) \cite{harris1990,fleis}, optical systems presenting a negative index of refraction \cite{veselago1968,smith2000}, cloaking devices \cite{meta}, one-way propagation and optical isolators \cite{jalas,pereira2014,jonas2016}, and light-trapping devices \cite{tsakmakidis2007,smolyaninov2010}, among many others. Regarding EIT, an experimental demonstration of the effect was reported in 1991, where an optically thick medium was shown to become transparent by means of a destructive interference caused by the application of an external electromagnetic field~\cite{Boller}.  More recently an EIT-like phenomenon was observed in the emission spectrum of a system of two resonant meta-atoms in a cavity as a response to an incident microwave~\cite{Hu}. See also the model and the experimental confirmation of a radiating two-oscillator model exhibiting EIT and absorption in metamaterials \cite{tassin2012}. There are several theoretical models examining EIT manifestations. For instance, it was shown  \cite{Zhang} that a plasmonic molecule exhibits an electromagnetic behavior that resembles the EIT in atomic systems, while Ref. \cite{tassin2009} shows that metamaterials in which electromagnetic radiation interacts resonantly with mesoscopic oscillators may display EIT. Furthermore, transparency can be controlled by changing some properties of the metamaterial \cite{kurter2011,stewart2013}. We should also mention the class of material media, commonly reported as {\em smart glasses}, whose optical properties can continuously run from opacity to transparency and back according to an externally controllable field imposed to the medium \cite{sm-glass,yasir,he}.

The aim of this work is to provide, for a given class of dielectric media and restricted to the limit of applicability of geometrical optics subjected to external electromagnetic fields, a theoretical description of the electromagnetically tunable optical opacity-to-transparency phase transition, while still relying upon the basic phenomenological properties of the medium.

This paper is summarized as follows. The next section presents the derivation of the Fresnel equation for light propagation in a non dispersive nonlinear medium by using the Hadamard method for field discontinuities. From the requirement of the existence of non-trivial solutions for the Fresnel equation, Sec.~\ref{suff-opac} deals with the geometric properties (speed and polarization) of the light rays, and establishes sufficient conditions for the occurrence of opacity in a non-magnetic medium. Furthermore, in this section we present and briefly discuss the effective metric that corresponds to the class of optical systems here considered. The standard description of electrically induced birefringence, by either linear Pockels or quadratic Kerr models, is but a mathematically simple account of electromagnetically tunable optical properties of the material medium. Such simplicity contrasts itself with the intrinsic sophistication of EIT phenomenon, thus suggesting the use of a more elaborate dependence on the field strength, as it was chosen in Secs.\ \ref{models} and \ref{magtc} below.
In Sec.~\ref{models}, simple models for the permittivity of the medium are investigated, and the way to control the opacity by means of an externally applied electric field is given. A very similar effect can also be obtained in cases for which the medium reacts nonlinearly to the presence of an external magnetic field, as shown in Sec.~\ref{magtc}. Conclusions and final remarks are given in Sec.~\ref{conclusion}. For the sake of completeness, the appendix determines the possible Jordan canonical forms for the Fresnel matrix.

\section{The light propagation in nonlinear optics}\label{math}
Let us consider Maxwell's equations inside a material medium in Euclidean space
\begin{eqnarray}
\p_iD_i&=&\rho,\label{max_diel1}\\[2ex]
\eta_{ijl}\p_jE_l&=&-\p_t B_i,\label{max_diel2}\\[2ex]
\p_iB_i&=&0,\label{max_diel3}\\[2ex]
\eta_{ijl}\p_jH_l&=& J_i + \p_t D_i.\label{max_diel4}
\end{eqnarray}
We are assuming the convention that Latin indices $i,j,l,\ldots...$ run from $1$ to $3$ and repeated indices are summed upon. We denote $\p_i=\p/\p x^i$, $\p_t=\p/\p t$, and $\eta_{ijl}$ is the completely skew-symmetric Levi-Civita symbol with $\eta_{123}=1$. Also, $E_i$ and $B_i$ stand for the electromagnetic field strengths, while $D_i$ and $H_i$ describe the corresponding field excitations (despite the generally accepted convention of term \(H_i\) as the magnetic field and \(B_i\) as the magnetic induction). Since nonlinear media are taken into account, the linear constitutive relations should be replaced by nonlinear ones,
\begin{eqnarray}
D_i=\epsilon_0E_i + P_i,&\quad\longrightarrow\quad& D_i = \epsilon_{ij}  E_j,\label{cons-rel1}\\[2ex]
H_i=\frac{1}{\mu_0}B_i-M_i,&\quad\longrightarrow\quad& H_i= \mu^{{}_{-1}}_{ij} B_j,\label{cons-rel2}
\end{eqnarray}
where $P_i$ is the polarization, $M_i$ is the magnetization, and $\epsilon_0$ and $\mu_0$ are the vacuum permittivity and permeability, respectively. It is generally assumed that both the phenomenological coefficients of permittivity  $\epsilon_{ij}$ and  inverse permeability $\mu^{{}_{-1}}_{ij}$ may be dependent upon the field strengths $E_i$ and $B_i$.

In the limit of geometric optics, $E_i$ and $B_i$, as well as the charge distribution $\rho$ and its flow $J_i$, are all assumed to be continuous through the wavefront $\Sigma_t(x_i)= \mbox{const.}$, for any given instant of time \(t\), but with a possibly nonzero finite ``step'' in their derivatives, as determined by the Hadamard method~\cite{hadamard} (see Ref.~\cite{dante} for a relativistic formulation of its application in geometrical optics)
\begin{eqnarray}
&\left[\p_t E_i\right]_{\Sigma_t}=-\omega\, e_i, \quad \left[\p_t B_i\right]_{\Sigma_t}=-\omega\, b_i,\label{h_omega}\\[1ex]
&\left[\p_i E_i\right]_{\Sigma_t}=e_i\,k_i, \quad \left[\p_i B_i\right]_{\Sigma_t}=b_i\,k_i,\label{h_inn}\\[1ex]
&\left[\eta_{ijl}\p_j E_l\right]_{\Sigma_t} = \eta_{ijl}k_j e_l\label{b},\quad \left[\eta_{ijl}\p_j B_l\right]_{\Sigma_t} = \eta_{ijl}k_j b_l\label{h_ext},
\end{eqnarray}
where $\omega$ is the wave frequency and $k_i$ is the wave vector. The vectors $e_i$ and $b_i$ describe the polarization modes of the electric and magnetic components of the light rays, respectively. We also use the symbol
\begin{equation}
\left[f\right]_{\Sigma_t}(p)=\lim_{\delta\to0^+}[f(p_+)-f(p_-)]
\end{equation}
to indicate how the step of an arbitrary function $f(p)$ through the surface $\Sigma_t$ is evaluated, where $p\in\Sigma_t$ and $p_{\pm}\in\Sigma_{t\pm\delta}$ are such that \(\lim p_{\pm}=p\) for \(\delta\to0^+\).

The above procedure, when applied to the Maxwell Eqs.\ (\ref{max_diel1})--(\ref{max_diel4}), yields \cite{local} a linearly polarized wave $\omega b_i =\eta_{ijl}k_j e_l$ whose electric polarization $e_i$ satisfies the eigenvalue problem
\begin{equation}
\label{eig_val}
Z_{ij} e_j=0.
\end{equation}
The components of the Fresnel matrix $Z_{ij}$ are given by \cite{goulart2008,pereira2014}
\begin{eqnarray}
Z_{ij}&=&\frac{1}{\omega}\left(\frac{\partial \epsilon_{ik}}{\partial B_l}\eta_{lmj}E_k + \frac{\partial \mu^{{}_{-1}}_{lk}}{\partial E_j}\eta_{iml}B_k\right)k_m
\nonumber \\
&&- \frac{1}{\omega^2}\eta_{ilm}\eta_{jqp}A_{mp}k_l k_q + C_{ij},
\label{fresnel}
\end{eqnarray}
where the auxiliary matrices $C_{ij}$ and $A_{ij}$ are defined as
\begin{eqnarray}
C_{ij}&=&\epsilon_{ij}+\frac{\partial \epsilon_{ik}}{\partial E_j}E_k,
\label{c_ij}
\\
A_{ij}&=&\mu^{{}_{-1}}_{ij}+\frac{\partial \mu^{{}_{-1}}_{ik}}{\partial B_j}B_k.
\label{h_ij}
\end{eqnarray}
The existence of nontrivial solutions for the eigenvalue problem given by Eq.\ (\ref {eig_val}) is equivalent to the requirement
\begin{equation}
\det[Z_{ij}]=0.
\label{fresnel_cond}
\end{equation}
This equation furnishes the dispersion relation associated with the light rays.

Up to now, the vectors $E_i$ and $B_i$ represent the total field, which should be understood as the composition of an external field \(E_i^{c}\) plus a wave field \(E_i^{\omega}\). Hereafter, we will assume that the fields associated with the propagating waves $E_i^{\omega}$ are much weaker than the controllable fields $E_i^{c}$, which can be produced by a distribution of sources in the medium or by means of external pumping fields. Thus, $E_i = E_i^{\omega} + E_i^{c} \approx  E_i^{c}$. For the sake of simplicity, we still make use of the same notation $E_i$ and $B_i$ for the components of the electric and magnetic fields, which for all practical purposes nearly coincide with the components of the external fields.

\section{Conditions for opacity in nonmagnetic systems}
\label{suff-opac}
Most of the material media we are able to optically operate with are such that they respond linearly to external magnetic fields. Therefore, we are allowed to simplify the model by assuming nonmagnetic systems such that $\epsilon_{ij}=\epsilon_{ij}(E_l)$ and $\mu^{{}_{-1}}_{ij}=(1/\mu)\delta_{ij}$, where $\mu$ is a constant permeability and $\delta_{ij}$ is the Kronecker delta, which is either $1$ for $i=j$ or $0$ otherwise. In this case, the Fresnel matrix given by Eq.~(\ref{fresnel}) reduces to
\begin{equation}
\label{red_fresnel}
Z_{ij} = C_{ij}-\frac{1}{\mu v^2}\Pi_{ij},
\end{equation}
where $v=\omega/k$ is the phase velocity. The wave vector can be written as $k_i=k\,\hat{k}_i$, with norm $k=\sqrt{k_i k_i}$, where $\hat k_i$ is the unit vector which points along the direction of the wave vector. Further, we define the projector on the plane orthogonal to the wave vector direction \(\hat k_i\) as $\Pi_{ij}=\delta_{ij}-\hat{k}_i\hat{k}_j$.

Straightforward calculations show that Eq.\ (\ref{fresnel_cond}) can be written as a polynomial equation for the phase velocity as
\begin{equation}
av^4+bv^2+d=0,
\label{eq-vp}
\end{equation}
where
\begin{eqnarray}
a&=&\det[C_{ij}],\label{coef-vp1}\\[1ex]
b&=&\frac{1}{\mu}\,\hat{k}_i\,[C_{il}C_{lj}-C_{ll}C_{ij}]\,\hat{k}_j,\label{coef-vp2}\\[1ex]
d&=&\frac{1}{\mu^2}\,\hat{k}_i\,C_{ij}\,\hat{k}_j.\label{coef-vp3}
\end{eqnarray}

The non-negative roots of Eq.\ (\ref{eq-vp}) are given by
\begin{equation}
v^{\pm}=\sqrt{-\frac{b}{2a}\left(1\pm \sqrt{1-\frac{4ad}{b^2}}\right)}.
\label{eq-vp-sol}
\end{equation}
Two possible cases of opacity inside the material medium are then identified from this equation, in terms of the discriminant \(\Delta=b^2-4ad\) of Eq.~(\ref{eq-vp}): either $\Delta<0$ or else $0\leq\Delta/b^2\leq 1$ with $b/a>0$. Note that such criteria for opacity is intended to determine algebraic relations for $C_{ij}$, since $\hat{k}_i$ is an arbitrary unit vector.

The algebraic criteria to verify whether the material under consideration admits a regime of opacity turns out to be useful when we already know the dielectric coefficients of such a medium. Otherwise, we have to solve partial differential equations for the coefficients $\epsilon_{ij}$ in order to set up the properties of the medium.
This latter approach shall be explicitly adopted in Sec.~\ref{models} in order to determine some specific models exhibiting the opacity phenomenon.

\subsection{Polarization}
The nonzero eigenvectors associated to the eigenvalue problem (\ref{eig_val}), namely the polarization vector $e_i$, can be found by considering a basis formed by any suitable choice of linearly independent vectors $\{\hat{u}^{(A)}_{\,i}: A=1,2,3\}$ with the index \(A\) labeling the three different vectors. We choose%
\footnote{Mathematically, however, $\hat k_i$ and $\hat E_i$ could be parallel to one another, a situation which would prevent them to be both included as basis vectors. Notwithstanding, we can rely on the continuity of Maxwell theory in order to {\it define} the optical behavior for such degenerate situation of \(\hat k_i\parallel\hat E_i\) as being given by that of the limit of \(|\eta_{pqr}\hat{k}_q\hat{E}_r|\) approaching zero without being equal to it.}
 the three unit vectors $\hat u^{(1)}_{\,i}=\hat{k}_i$, $\hat{u}^{(2)}_{\,i}=\hat{E}_i$, and $\hat{u}^{(3)}_{\,i}=\eta_{ijl}\hat{k}_j\hat{E}_l/|\eta_{pqr}\hat{k}_q\hat{E}_r|$, where $\hat{E}_i=E_i/E$ is the unit vector directed along the external electric field $E_i$, in which $E=|E_i|$ is the magnitude of the electric field, and $|X_i|$ denotes the Euclidean norm of any given vector \(X_i\).
The decomposition of $e_i$ in terms of the basis \(\{\hat{k}_i,\,\hat{E}_i,\,\eta_{ijl}\hat{k}_j\hat{E}_l/|\eta_{pqr}\hat{k}_q\hat{E}_r|\}\) is
\begin{equation}
e_i=a_{A}\,\hat u^{(A)}_i,
\end{equation}
where $a_A$ are constants. For the phase velocity \(v\) of the light ray being given by Eq.~(\ref{eq-vp-sol}), the coefficients $a_A$ of the polarization vector \(e_i\) are determined as
\begin{equation}
\label{eq_ai_cas0}
\begin{array}{l@{\hspace{.3em}}l}
a_1=&\left\{\left[\hat E_sC_{st}\hat E_t-\frac{1-(\hat E_n\hat k_n)^2}{\mu v^2}\right]\hat k_i-(\hat k_sC_{st}\hat E_t)\hat E_i\right\}
\\[1ex]&\times C_{ij}\frac{\eta_{jlm}\hat k_l\hat E_m}{|\eta_{pqr}\hat k_q\hat E_r|},\\[1ex]
a_2=&(\hat k_i\hat E_s-\hat k_s\hat E_i)C_{st}\hat E_tC_{ij}\frac{\eta_{jlm}\hat k_l\hat E_m}{|\eta_{pqr}\hat k_q\hat E_r|},\\[1ex]
a_3=&\hat k_sC_{st}\hat E_iC_{ij}(\hat k_t\hat E_j+\hat k_j\hat E_t)+\hat k_sC_{st}\hat k_t\frac{1-(\hat E_n\hat k_n)^2}{\mu v^2},
\end{array}
\end{equation}
up to an arbitrary global normalization.

\subsection{Remarks on optical analog models}
A simple manipulation of Eq.\ (\ref{eq-vp}) allows us to cast it as \(g^{\mu\nu}k_\mu k_\nu=0\), from which we can read out an effective optical geometry \cite{edu2012,jonas} describing the dispersion relation associated with the four-dimensional wave vector $k_{\mu}\doteq (\omega,k_i)$, with Greek indices running from $0$ to $3$. Here Minkowski background metric, which in Cartesian coordinates takes the form  $[\eta_{\mu\nu}] = {\rm diag}(-1,1,1,1)$, associates tensors with lower and upper indices.  In terms of the quantities defined before, this effective metric can be written as \cite{teodoro}
\begin{equation}
\label{eff_metric}
g^{\mu\nu}=-\mu a\,V^{\mu}V^{\nu}+\left[C^\alpha{}_\alpha-\frac{1}{\mu (v^{\pm})^2}\right]C^{\mu\nu}-C^{\mu\alpha}C_{\alpha}{}^{\nu},
\end{equation}
where $V^{\mu}=\delta^{\mu}{}_{0}$ and $C_{\mu\nu}$ is defined such that for either $\mu$ or $\nu$ equal to zero, then $C_{\mu\nu}=0$; otherwise, it coincides with $C_{ij}$. It is straightforward to show that for the particular case of vacuum, where the dielectric coefficients are just $\epsilon_0$ and $\mu_0$, the optical metric $g^{\mu\nu}$ reduces to the Minkowski metric $\eta^{\mu\nu}$. One important point in this description is that the integral curves of the wave vector $k_\mu$ in the flat Minkowski spacetime correspond to null geodesic curves in the fictitious spacetime whose metric is given by the effective geometry \cite{delorenci2002,delorenci2002b}.

The usefulness of the effective metric description relies on the fact that the light propagation inside an optical medium described by dielectric coefficients in a flat spacetime is mathematically equivalent to the light propagation in a curved spacetime. Generically speaking, a curved spacetime is an exact solution of general relativity for a given source of energy. Therefore, the coefficients of the effective geometry could be compared with the coefficients of the curved spacetime geometry and, as far as only kinematic aspects of such propagation are considered, the same sort of phenomena predicted in gravitational systems could be investigated in the context of optics in material media. Possible applications of this analogy include the tests of bending of light and the construction of cosmological models in terrestrial laboratories. Its theoretical relevance is also linked with the possibility of measurement of tiny quantum phenomena such as Hawking radiation, or even testing quantum gravity predictions, as is discussed in the literature \cite{barcelo2005}.

\section{Simple models for smart glasses}\label{models}
We provide here some simple examples in which the material media present opacity for a given range of the electromagnetic field and the incident angle of the light rays. One should notice that the derivation of the results below is valid for any material satisfying the sufficient requirements stated in the previous section---just after Eq.~(\ref{eq-vp-sol})---and not only the smart glasses already known in the literature. Notwithstanding, we shall keep calling those media as smart or switchable glasses, because we believe that these words appropriately express the idea behind the formalism developed here.

Let us assume a magnetically linear medium whose permittivity is $\epsilon_{ij}=\epsilon_{ij}(E)$ dependent only on the magnitude \(E\) of the electric field. Suppose that the external electric field lies on the $(x,y)$-plane, that is, $[E_i]=(E\cos\phi,E\sin\phi,0)$, with $\phi\in(0,2\pi)$ and let the unit wave vector be $[\hat{k}_i]=(k_x,k_y,k_z)$ with $k_x^2+k_y^2+k_z^2=1$. Suppose that the permittivity tensor is diagonal, $[\epsilon_{ij}]= \mbox{diag}(\epsilon_1,\epsilon_2,\epsilon_3)$. The auxiliary matrix $[C_{ij}]$ then reads
\begin{equation}
\label{c-sg}
[C_{ij}]=\left[
\begin{array}{ccc}
\epsilon_1+ \epsilon_{1}'\cos^2\phi&\epsilon_{1}'\cos\phi\sin\phi&0\\[1ex]
\epsilon_{2}'\cos\phi\sin\phi&\epsilon_2 + \epsilon_{2}'\sin^2\phi&0\\[1ex]
0&0&\epsilon_3
\end{array}\right],
\end{equation}
where we denote $X'=E (dX/dE)$. The coefficients of the polynomial equation (\ref{eq-vp}) for the phase velocity are obtained from
\begin{equation}
\label{coeff_ph_vel}
\begin{array}{rl}
a=&\epsilon_3(\epsilon_1\epsilon_2+ \epsilon_1\epsilon_2'\sin^2\phi+ \epsilon_2\epsilon_1'\cos^2\phi),\\[1ex]
\mu b=& - \frac{1}{2}(\epsilon_1' + \epsilon_2') \epsilon_3 k_x k_y \sin(2\phi) \\[1ex]
&+(\epsilon_1 - \epsilon_2 + \epsilon_1'\cos^2\phi - \epsilon_2'\sin^2\phi) \epsilon_3 k_y^2\\[1ex]
&+[(\epsilon_1 - \epsilon_3)(\epsilon_2+\epsilon_2'\sin^2\phi) +\epsilon_2\epsilon_1'\cos^2\phi]k_z^2 \\[1ex]
&- \epsilon_1(\epsilon_2+\epsilon_3) - (\epsilon_2+\epsilon_3)\epsilon_1'\cos^2\phi- \epsilon_1\epsilon_2'\sin^2\phi,\\[1ex]
\mu^2d=&\frac{1}{2}(\epsilon_1' + \epsilon_2') k_x k_y \sin(2\phi)+ \epsilon_3k_z^2 \\[1ex]
&+(\epsilon_2 - \epsilon_1 + \epsilon_2'\sin^2\phi - \epsilon_1'\cos^2\phi) k_y^2\\[1ex]
&+(\epsilon_1+\epsilon_1'\cos^2\phi)(1 - k_z^2).
\end{array}
\end{equation}
The decomposition of the unit wave vector in spherical coordinates is $[\hat{k}_i]=(\sin\theta\cos\varphi,\sin\theta\sin\varphi,\cos\theta)$, where $\theta\in(0,\pi)$ and $\varphi\in(0,2\pi)$.
Note that the angles \(\phi\) and \(\varphi\) are physically nonequivalent variables.

In what follows, we particularize our investigation to the cases where $\epsilon_1=\epsilon(1+f)$, $\epsilon_2=\epsilon(1-f)$, and $\epsilon_3=\epsilon$, where $f=f(E)$ and $\epsilon$ is a constant, in order to determine some simple models exhibiting controlled opacity to transparency phase transitions.

\subsection{A simple model for electrically switchable media}
\label{elect}
Let us choose $f(E)$ such that the phase velocities, given by Eq.\ (\ref{eq-vp-sol}), exhibits azimuthal symmetry with respect to the wave vector direction. In this case, the coefficients $a$, $b$, and $d$, given by Eq. (\ref{coeff_ph_vel}), will not depend on the components $k_x$ and $k_y$ of the wave vector; i.e., they will not depend on the azimuthal angle $\varphi$. Such requirement is achieved by setting $f'+2f=0$, which leads to $f=(E_{{}_0}/E)^{2}$, where $E_{{}_0}$ is an integration constant.%
\footnote{We recall that the model being proposed here emphasizes mathematical simplicity, and possibly disregards other aspects which may prove relevant elsewhere. For instance, the regularity of Taylor expansions could be recovered by a function which roughly resembles a Planckian distribution \(f(x)=1/\{x^3[\exp(1/x)-1]\}\) with \(x=E/E_\omega\), which is globally regular and scales as \(1/x^2\) in the \(x\gg1\) regime as required.}
The adopted regime of wave propagation we are assuming in this paper requires that the intensity of the wave field, hereafter denoted by $E_\omega$, has to be small when compared to the external field $E$. In order to ensure the validity of this regime, we set $E_{{}_0} = E_\omega$, which implies that $f\ll1$.

For this particular choice of $f$, the discriminant \(\Delta\) of Eq.\ (\ref{eq-vp}) reduces to
\begin{equation}
\Delta=\frac{\epsilon^4}{\tilde E^4\mu^2} \big[h\,\sin^4\theta - \sin(2\phi) (1+\cos^2\theta)^2\big],
\end{equation}
where $h=1 - (2/\tilde E^2)\cos(2\phi) + 1/\tilde E^4$ is such that \(h\ge0\), and $\tilde E=E/E_\omega$. Note that small values of $\theta$ imply $\Delta<0$ whenever $\sin(2\phi)>0$. For $\phi=n\pi/2$ with $n\in\mathbb{Z}$, we have $\Delta\geq0$. Therefore, $\Delta=0$ and $\sin(2\phi)\neq0$ leads to
\begin{equation}
\label{cos_delta}
\cos^2\theta = \frac{1 - \sqrt{\fracc{\sin(2\phi)}{h}}}{1 + \sqrt{\fracc{\sin(2\phi)}{h}}}.
\end{equation}

From the above analysis, it follows that $\phi$ have to be in the interval $(0,\pi/2)$ in order for the opacity phenomenon to exist. Such opacity is most intense in the middle of the interval: $\bar\phi=\pi/4$. In this case, Eq.~(\ref{cos_delta}) implies that
\begin{equation}
\cos^2\theta=\frac{1-\sqrt{\frac{\tilde E^4}{1+\tilde E^4}}}{1+\sqrt{\frac{\tilde E^4}{1+\tilde E^4}}}.
\end{equation}
As $\tilde E$ is larger, $\cos^2\theta$ becomes smaller and, therefore, the range of the polar angle \(\theta\) for which $\Delta<0$ is larger.

\begin{figure}[ht]
\begin{center}
\includegraphics[width=8cm]{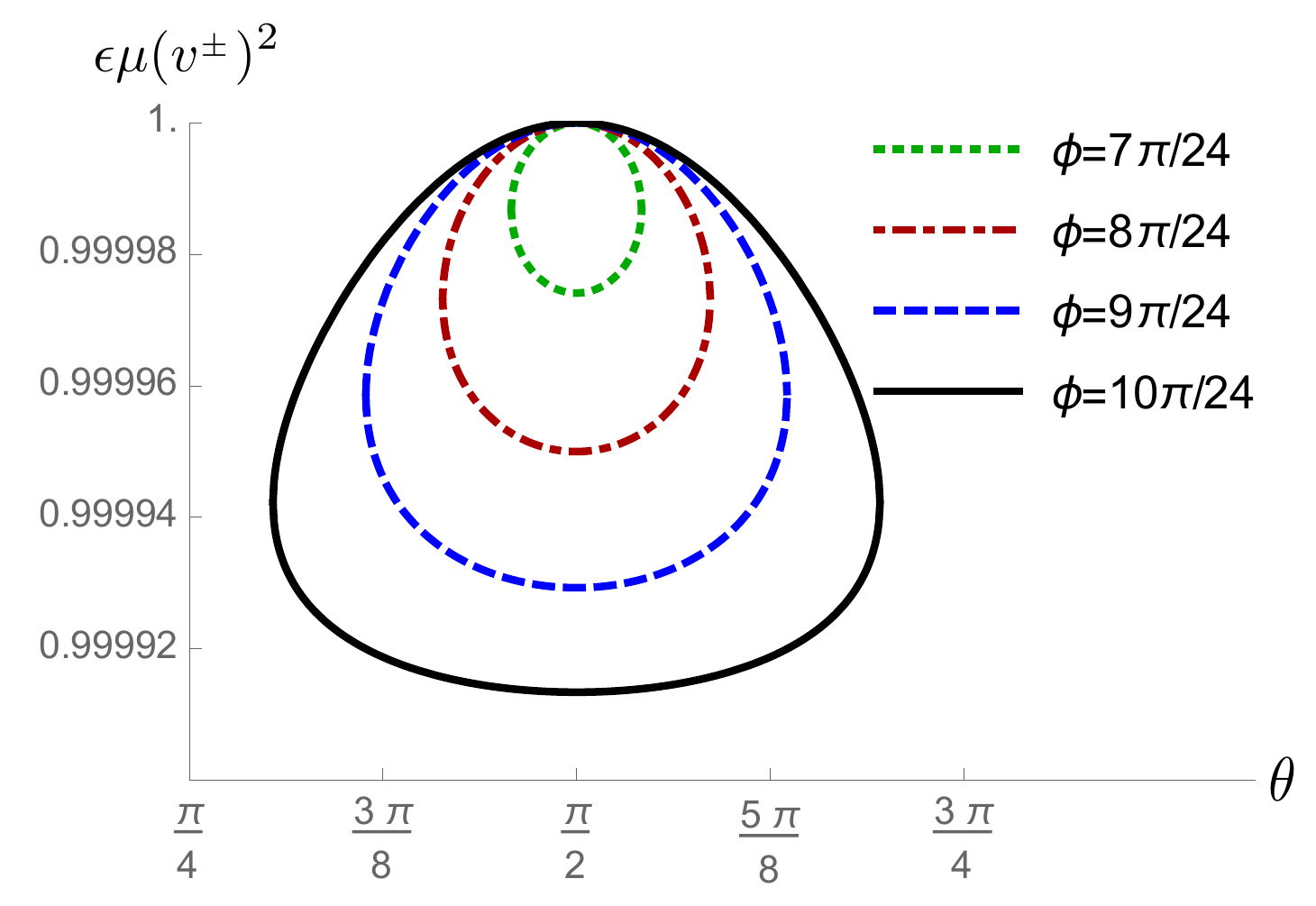}
\end{center}
\caption{Plots of $\epsilon\mu (v^\pm)^2$ as function of $\theta$ for the model given by $f=(E_{{}_\omega}/E)^{2}$. It was set  \(E= 100 E_\omega\).}
\label{figa}
\end{figure}

The above model is numerically studied in Fig.\ \ref{figa} where $\epsilon\mu (v^\pm)^2$ as function of $\theta$ is depicted for some representative value of $\phi$. Notice that as $\phi$ gets larger values compared to $\bar\phi$, the window of transparency enlarges, as anticipated.
As one can confirm by inspecting the coefficients $a$, $b$, and $d$ in Eq.~(\ref{coeff_ph_vel}), the phase velocity is independent of the polar direction $\varphi$ of the wave vector in this model.

For the maximum of opacity ($\phi=\bar\phi$), the polarization vectors are determined by Eqs.\ (\ref{eq_ai_cas0}) as
\begin{eqnarray}
\begin{array}{l}
a_3 = \frac{[\sqrt{2}a_1\tilde v^2+a_2\sin\theta(\cos\varphi+\sin\varphi)]\sqrt{\beta_0}\cot\theta}{\sqrt{2}(\sin\varphi-\cos\varphi)(\tilde v^2-1)},
\\[1.5ex]
\frac{a_1}{a_2} = \frac{2-(f+1)\tilde v^2 + (1+\beta_1\cos^2\theta)f - \beta_1\left(1 -\frac{\tan^2\theta \sin\varphi}{\beta_2}\right)\left(1-\frac{1}{\tilde v^2}\right)}{\sqrt{2}\left\{\left[\beta_3+ \beta_2(1-f)\right] \tilde v^2 - \beta_3 +\frac{\beta_2}{\sin\theta}\right\}},
\end{array}
\end{eqnarray}
in which
\begin{eqnarray}
\label{eq_coef_v}
\begin{array}{l}
\left[\frac{\mu\epsilon b}{a}+ \frac{\beta_0\sin^2\theta f^2 + 2\cos^2\theta f + 2(\beta_0-\cos^2\theta)}{\beta_0(f^2 + 1)}\right]\tilde v^2
\\[1.5ex]
+\frac{\mu^2\epsilon^2 d}{a}+\frac{\sin^2\theta[(\cos(2\varphi) - \sin(2\varphi)-1)\cos^2(\theta) f - 1 + \sin(2\varphi)]}{\beta_0(f^2 + 1)} = 0,
\end{array}
\end{eqnarray}
where $\tilde v^2=\mu\epsilon (v^{\pm})^2$---with the phase velocities $v^{\pm}$ being given by Eq.\ (\ref{eq-vp-sol}) restricted to this particular case. We also define some auxiliary parameters:  $\beta_0=1+\cos^2\theta-\sin^2\theta\sin(2\varphi)$, $\beta_1=(\sin\varphi +\cos\varphi )/(\sin\varphi -\cos\varphi )$, $\beta_2=\cos^2\theta/(\sin\varphi -\cos\varphi )$, and $\beta_3=\sin\theta(f\cos\varphi+\sin\varphi)$. Notice that Eq.\ (\ref{eq_coef_v}) involves only the angular coordinates \(\theta,\,\varphi\) of the wave vector, that is, it establishes a relationship between the angles in the form $\theta=\theta(\varphi)$ or vice versa, since the dimensionless phase velocity \(\tilde v\) was already determined from Eq.~(\ref{eq-vp-sol}).

Finally, we compute the determinant of the effective metric $g\equiv\det{[g_{\mu\nu}]}$, given by Eq.\ (\ref{eff_metric}), as
\begin{equation}
\begin{array}{l@{\hspace{.2em}}c@{\hspace{.2em}}l}
\hspace*{-0.5em}
g&=& -\mu\epsilon(1 + f^2) \left[ \left(2 + f^2 - \frac{1}{\tilde v^2}\right)^2 +f^2\left(1-\frac{1}{\tilde v^2} \right)^2 \right]
\\&&\times\left(2-\frac{1}{\tilde v^2}\right).
\end{array}
\end{equation}
From this, it follows that the condition $g<0$ (the hyperbolicity condition for the propagation of the light ray) holds only if $\mu\epsilon(v^{\pm})^2>1/2$, in agreement with the regime of geometrical optics that we have assumed.

\subsection{Electrically switchable media with thin film polarizers}
There are other simple models that one may propose by taking into account a thin film polarizer on the interface between the dielectric and the empty space, so that one of the wave vector directions tangent to the $(x,y)$ plane will be absorbed by the polarizer, for example, $k_y=0$. In this case, two interesting profiles for $f(E)$ are discussed below.

\subsubsection{Power law models}
If we choose $f = (E/E_\omega)^{x}$ with $E>E_\omega$ and $E_\omega$ as the probe field, Eq.\ (\ref{eq-vp-sol}) gives the following expressions for the phase velocities
\begin{eqnarray}
\begin{array}{l}
\mu\epsilon(v^{\pm})^2=\fracc{[x+1-(\frac{x}{2}+1)\tilde E^{-x}]\sin^2\theta-2\tilde E^{-2x}}{2 (x + 1- \tilde E^{-2x})}
\\[2ex]
\times\left[1\pm \sqrt{1+4\fracc{\tilde E^{-x}(x+1- \tilde E^{-2x})[(\frac{x}{2}+1)\sin^2\theta + \tilde E^{-x}]}{\{[x+1-(\frac{x}{2}+1)\tilde E^{-x}]\sin^2\theta-2\tilde E^{-2x}\}^2}}\right].
\end{array}
\label{pl}
\end{eqnarray}
For small values of $\theta$, we obtain that the condition $\Delta<0$ requires $x<-1$, which particularly includes the case \(f\sim E^{-2}\)  analyzed in Sec. \ref{elect}. It should be remarked that the critical field can always be thought as the probe field one; therefore it is expected that power-law models have negative exponents in order that $f<1$.
\begin{figure}[ht]
\begin{center}
\includegraphics[width=8cm]{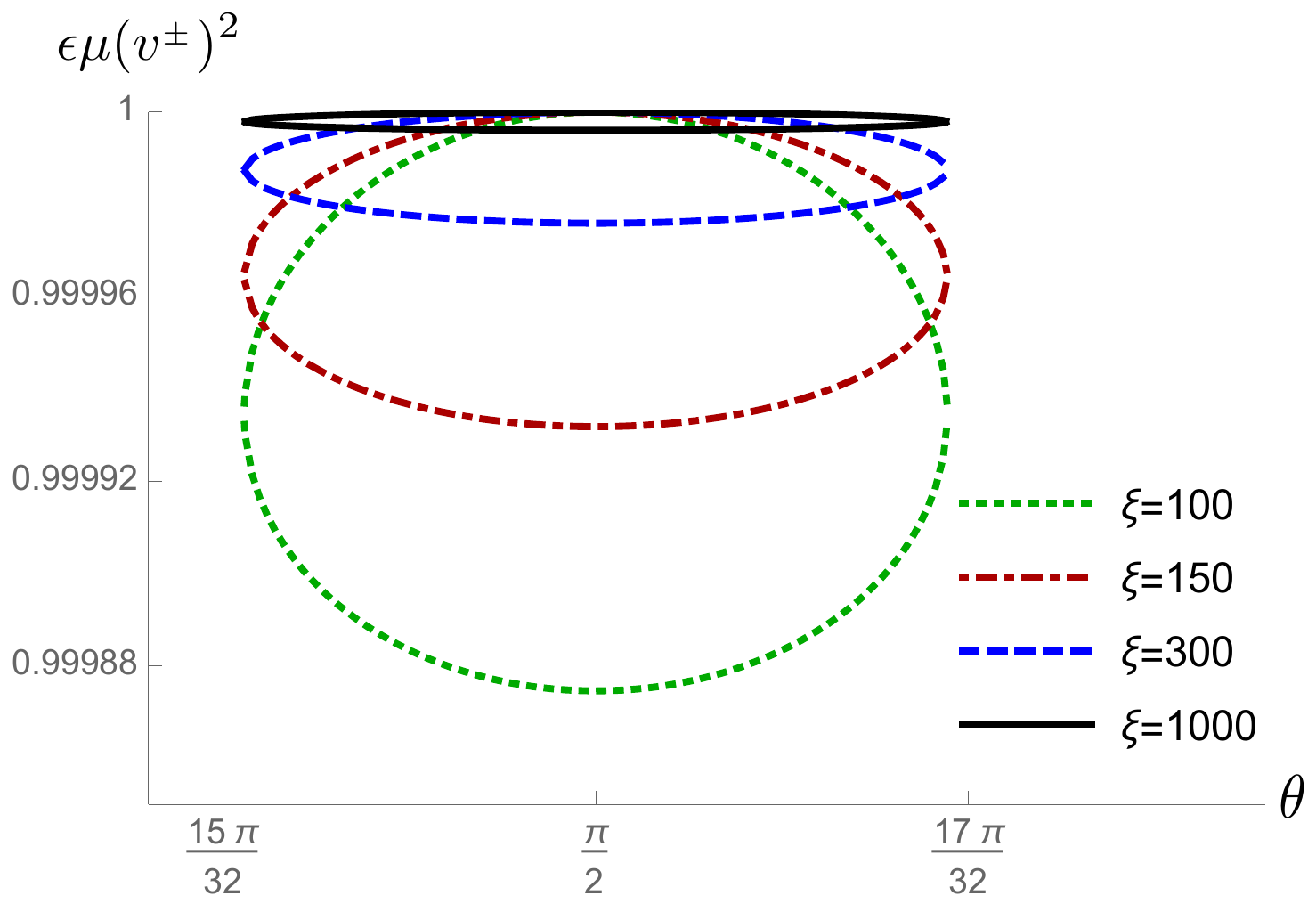}
\end{center}
\caption{Plots of $\epsilon\mu (v^\pm)^2$ as function of $\theta$ for the power-law model described by Eq.~(\ref{pl}) with $x=-1.05$. The solutions were set with \(E=\xi E_\omega\) for a few illustrative values of \(\xi\). Here $[\hat{k}_i]=(\sin\theta,0,\cos\theta)$ and  $[\hat E_i] = (1/2)\,(1,\sqrt{3},0)$.}
\label{figb}
\end{figure}
Figure~\ref{figb} depicts the behavior of the phase velocities, by means of $\epsilon\mu (v^\pm)^2$, described by Eq.~(\ref{pl}) for some representative values of the ratio $E/E_\omega$ when the specific model given by $x=-1.05$ is set. Notice that for this case, the medium exhibits transparency in a narrow and symmetric angular range of approximately $\pi/16$ rad around $\theta = \pi/2$; otherwise it is opaque. Furthermore, the width of the angular range for which transparency occurs does not depend on the magnitude of the electric field. The magnitude $E$ of the field affects only the magnitude \(v^\pm\) of the phase velocity.
This behavior, including the range of velocities and the widths of the transparency windows, is highly dependent on the specific value of the power $x$ which characterizes each possible model.

\subsubsection{Exponential models}
\begin{figure}[ht]
\begin{center}
\includegraphics[width=8cm]{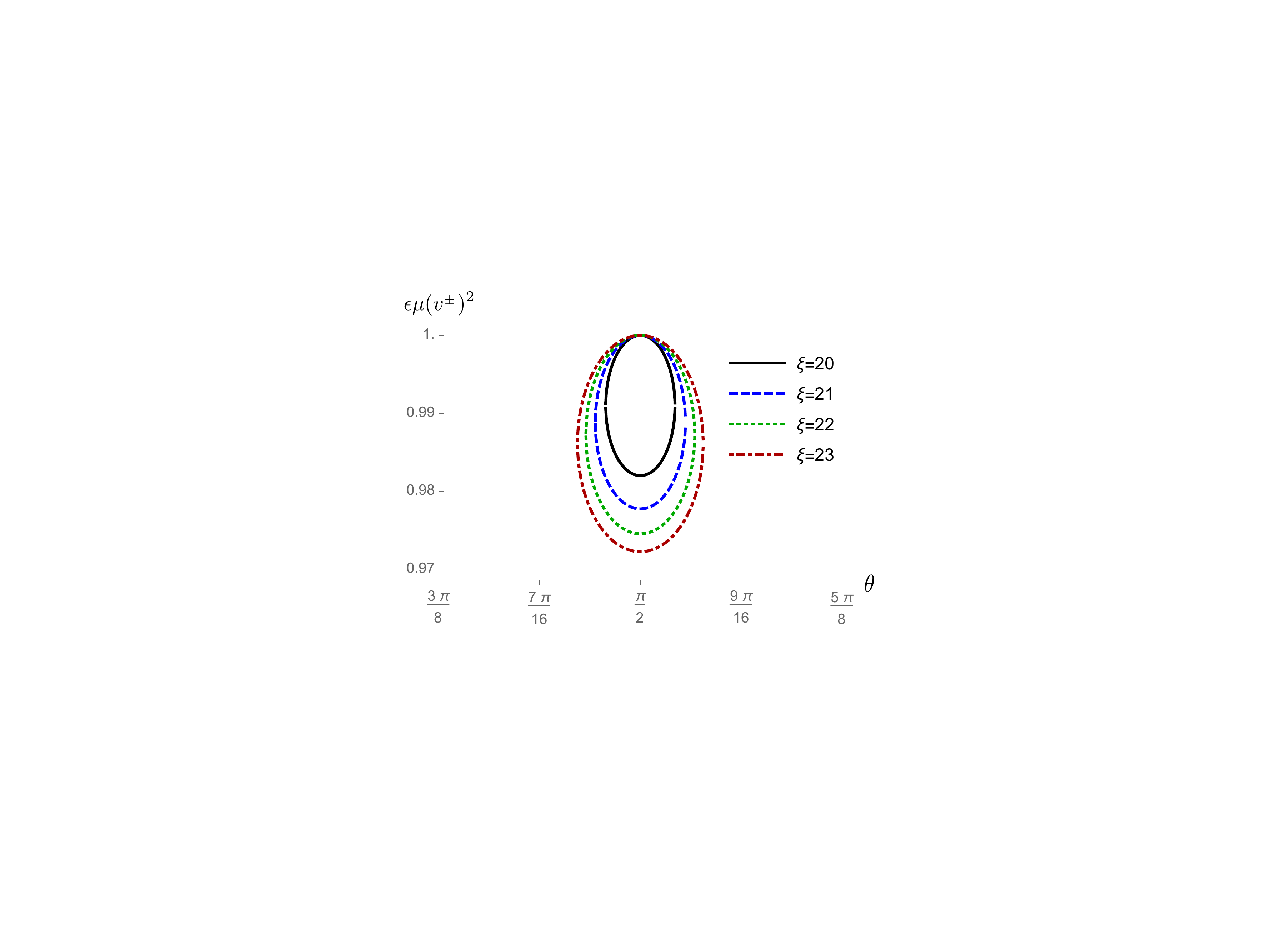}
\end{center}
\caption{Plots of $\epsilon\mu (v^\pm)^2$ as function of $\theta$ in the case of an exponential model with $\alpha = 0.1$. The solutions were set with \(E=\xi E_\omega\) for a few illustrative values of \(\xi\). Here  $[\hat E_i] = (1/\sqrt{2})\,(1,1,0)$.}
\label{fig2}
\end{figure}

Material media characterized by $f= \exp (-\alpha E/E_\omega)$, with $\alpha$ being a constant, also present opacity. Figure \ref{fig2} depicts the behavior of the phase velocities for some representative values of the ratio $\alpha E/E_\omega$. Again, we can set $E_\omega$ as the probe field, from which the limit of geometric optics requires $E>E_\omega$. Therefore, $f$ can be seen as a small correction to the dielectric permittivity of an isotropic medium.
This figure shows that the system exhibits transparency only in a narrow window  about $\theta = \pi/2$; otherwise it is opaque. The width of the transparency window for this case is dependent on the magnitude $E$ of the external electric field: A larger electric field opens a larger window of transparency.

\section{Opacity in magnetic media}
\label{magtc}
\subsection{Magnetically switchable media}\label{mag}
We may also take advantage of the possibly nonlinear behavior of the dielectric parameters in the magnetic fields. Let us consider a medium whose dielectric coefficients are \(\epsilon_{ij}=\epsilon\,\delta_{ij}\) and \(\mu^{{}_{-1}}_{ij} = (1/\mu)\mbox{diag} (1+m, 1-m,1)\) with $m=m(B)$ depending upon the magnitude of the magnetic field, while $\epsilon$ and $\mu$ are constant parameters. The dispersion relation stated in Eq.~(\ref{fresnel_cond}) yields in this situation a biquadratic equation for the phase velocity,
\begin{eqnarray}
\label{ph_vel_mag-0}
&&\mu(\hat{k}_2^2 A_{11}+\hat{k}_1^2 A_{22}) + \mu^2\hat{k}_3^{2}(A_{12}^2+A_{11}A_{22})
\nonumber\\
&& -\left[\hat{k}_1^2 +\hat{k}_2^2 + \mu(\hat{k}_2^2+\hat{k}_3^{2})A_{11} + \mu(\hat{k}_1^2+\hat{k}_3^{2})A_{22}\right] \mu\epsilon v^{2}
\nonumber\\
&&+(\mu\epsilon v^2)^2=0,
\end{eqnarray}
where we assumed the magnetic field points along the direction $[\hat B_i]=(\cos\phi,\sin\phi,0)$.

\begin{figure}[ht]
\begin{center}
\includegraphics[width=8cm]{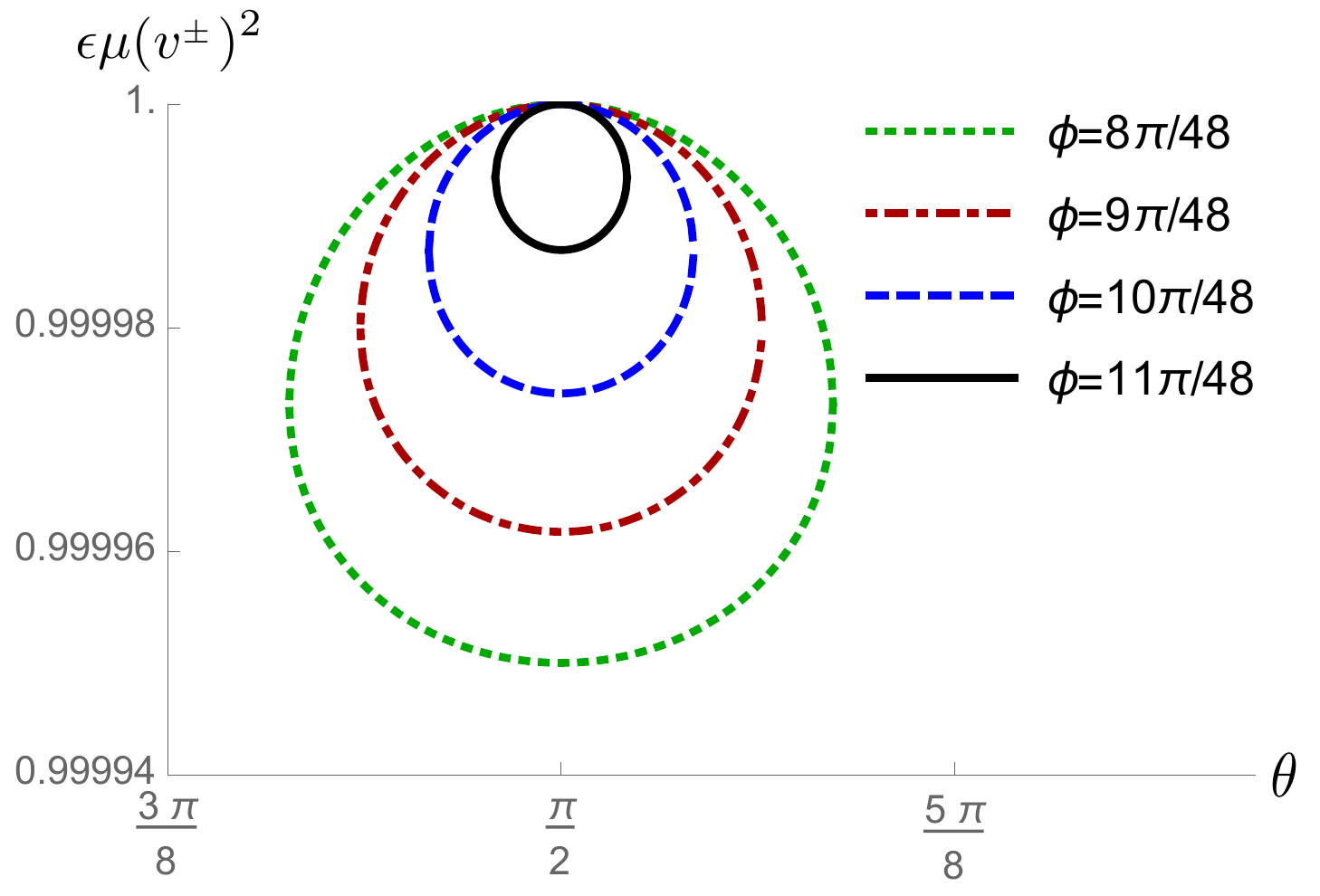}
\end{center}
\caption{Plots of $\epsilon\mu (v^\pm)^2$ as function of $\theta$ for the model given by $m=(B_\omega/B)^{2}$. Here  \(B= 100 B_\omega\).}
\label{fig_mag}
\end{figure}

The same steps presented in Sec.~\ref{elect} for the electric case now yield $m(B)=(B_{{}_0}/B)^2$, with $B_0$ a constant, and we end up with
\begin{eqnarray}
\label{ph_vel_mag}
(\mu\epsilon v^2)^2 -\left[\mu A_{11} (1+\hat k_3^{2}) + \mu A_{33} (1-\hat k_3^{2})\right] \mu\epsilon v^{2}
\nonumber\\
+\mu^2(A_{11}^{2}+A_{12}^{2})\hat k_{3}^{2}+\mu^2A_{11}A_{33} (1- \hat k_{3}^{2})=0,
\end{eqnarray}
where the components $A_{ij}$ are computed from the definition (\ref{h_ij}). The decomposition in spherical coordinates for the wave vector $[\hat{k}_i]=(\sin\theta\cos\varphi,\sin\theta\sin\varphi,\cos\theta)$ then yields the discriminant \(\Delta\) of Eq.~(\ref{ph_vel_mag}) in the form
\begin{equation}
\Delta=m(B)^2[\sin^2\theta-\sin^2(2\phi)(1+\cos^2\theta)^2].
\label{deltaD}
\end{equation}
We note that the occurrence of opacity (which is achieved from the requirement \(\Delta<0\)) is more easily obtained in this case as compared to the electric ones, since it does not depend on the intensity of the magnetic field but only on its direction. In particular, if $\phi$ is equal to either $0$ or $\pi/2$, then the medium is completely transparent (and  birefringence phenomena is expected to occur) for rays with \(0<\theta<\pi\); on the other hand, for $\phi=\pi/4$, the medium is completely opaque. Figure \ref{fig_mag} illustrates the behavior of the phase velocities for some representative values of $\phi$. Recall that the window of transparency diminishes as $\phi$ enlarges.

In this model, as the sign of $\Delta$  depends only on the direction ($\phi$) of the external magnetic field, for a given direction ($\theta$) of propagation, opacity-to-transparency transition could be controlled by using an apparatus built with movable permanent magnets. Such a hypothetical device would be safer to handle, as compared with the electrical cases discussed in the previous sections.

\subsection{Electromagnetically switchable media}\label{electromag}
For the sake of completeness, we present also the case concerning smart glasses generated by nonlinearities in both dielectric parameters, with an anisotropy in the electric sector.  We assume the permittivity coefficients of the medium as
\begin{equation}
[\epsilon_{ij}]=\epsilon(1-\lambda E^2)\,\mbox{diag}(\gamma,1,1),
\label{ep}
\end{equation}
 and the permeability coefficients
 \begin{equation}
 \mu^{{}_{-1}}_{ij}=(1/\mu)(1-\sigma B^2)\delta_{ij},
 \label{pe}
 \end{equation}
both with linear dependence upon the squared magnitude of the fields, where \(\epsilon,\mu,\lambda,\sigma,\gamma\) are positive constant parameters. Notice that the problem admits immediate generalization for metamaterials, with \(\epsilon\) and \(\mu\)  taking negative values. Most crystalline optical materials have dielectric coefficients presenting a similar mathematical form but satisfying the conditions $\lambda E^2 \ll1$ and $\sigma B^2\ll1$. This is not the regime being considered in the present work. The effects we analyze here require the acceptance of Eqs. (\ref{ep}) and (\ref{pe}) as exact functions of the fields.

The expression in Eq.~(\ref{fresnel}) is greatly simplified by imposing the magnitudes of the electric and magnetic fields to be chosen such that \(\lambda E^2=1/3=\sigma B^2\). For their angular orientations, we choose \([E_i/E]=(\sin\chi,\,0,\,\cos\chi)\) and \([B_i/B]=(0,\,0,\,1)\). The dispersion relation can then be given as a linear equation for \(v^2\),
\begin{eqnarray}
\mu\epsilon\,v^2\!\!&=\!\!&\big[4\sin^2\theta\left(\gamma\cos^2\chi\cos^2\varphi +\sin^2\varphi\right)+ 4\cos^2\theta\sin^2\chi
\nonumber\\
&& -(1+\gamma)\sin(2\theta) \sin(2\chi)\, \cos\varphi\big]
\left[4\gamma\cos^2\chi\tan^2\theta \right.
\nonumber\\
&&+ \left.4\sin^2\chi -2(1+\gamma)\tan\theta\,\sin(2\chi)\,\cos\varphi\right]^{-1},
\label{vEBg}
\end{eqnarray}
where $\theta$ and $\varphi$ represent, respectively, the polar and the azimuthal angles associated with the wave vector direction $[\hat{k}_i]=(\sin\theta\cos\varphi, \sin\theta\sin\varphi, \cos\theta)$.

A remarkable feature of this solution is the absence of birefringence. In fact, the chosen values for the electric and magnetic fields are such that the medium satisfies a sort of nonbirefringence condition \cite{teodoro}.

\begin{figure}
\begin{center}
\mbox{
\includegraphics[width=8.5cm]{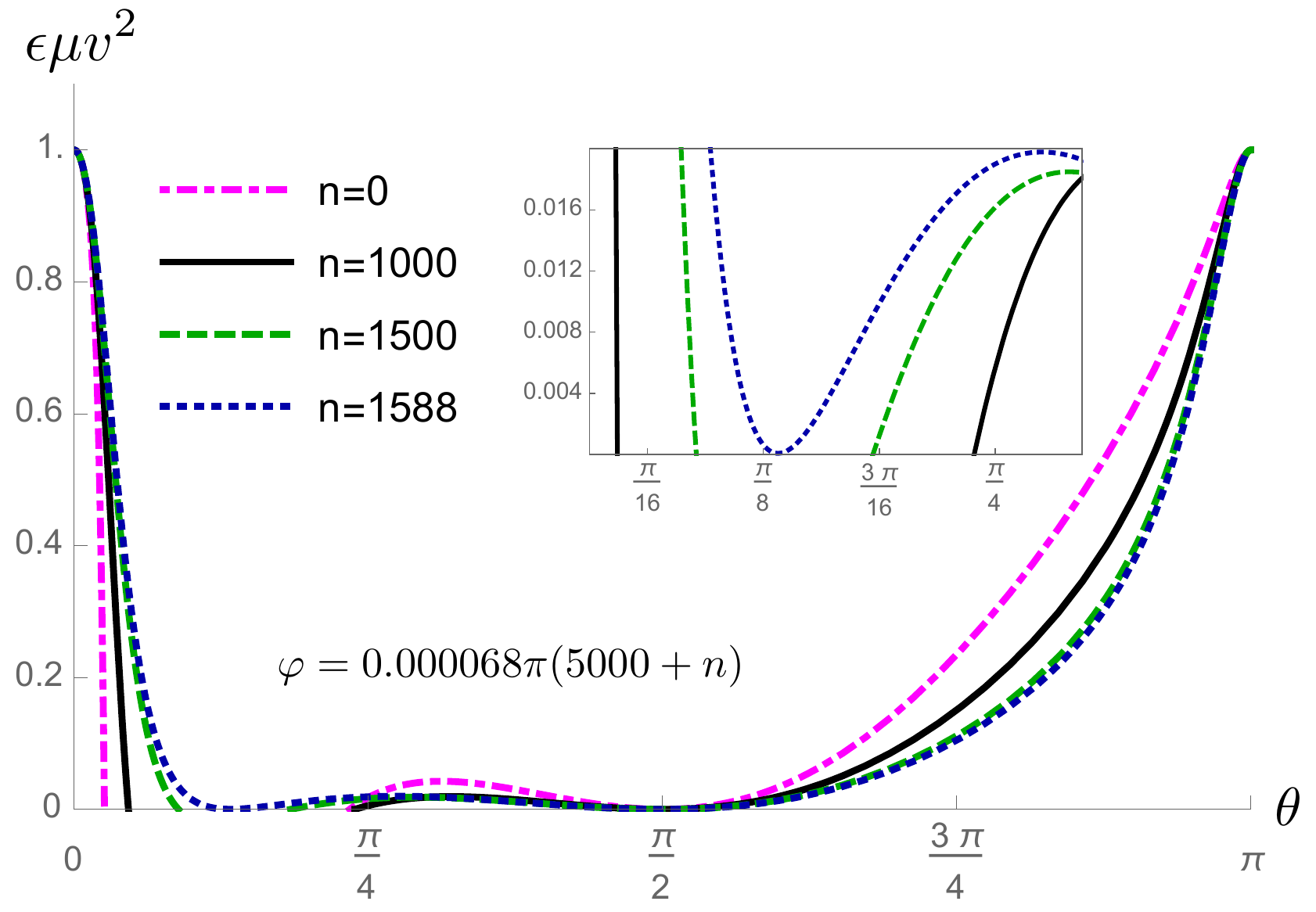}
}
\end{center}
\caption{Plots of \(\mu\epsilon\,v^2\) as a function of $\theta$ from Eq.~(\ref{vEB}) with \(\gamma=15\), for \(\varphi\) slightly larger than \(8\pi/25\). For \(\varphi<12\pi/25\) there exists a window of opacity near \(\theta=\pi/8\), as depicted in the inset frame for $n=1000, 1500, 1588$.
}
\label{fig3}
\end{figure}

In order to study a specific case, let us set \(\chi=\pi/6\). In this case we obtain that
\begin{eqnarray}
\mu\epsilon\,v^2&=&\cos^2\theta\,\big[
\cos^2\theta+\sin^2\theta(4\sin^2\varphi+3\gamma\,\cos^2\varphi)
\nonumber\\&&
-\frac{\sqrt3}{2}(1+\gamma)\sin\,(2\theta)\,\cos\,\varphi\big]\big[\cos^2\theta
\nonumber\\&&
+3\gamma\,\sin^2\theta
-\frac{\sqrt3}{2}(1+\gamma)\sin\,(2\theta)\cos\varphi\big]^{-1},
\label{vEB}
\end{eqnarray}
which shows that the system presents opacity for small values of  the angle \(\phi\) for a range of the angle \(\theta\), if \(\gamma>4/3\). This solution is depicted for \(\gamma=15\) in Fig.~\ref{fig3} for some illustrative values of $\varphi$.

The relevance of the two cases with the magnetic field lies in the fact that the propagation (with reference to the phase velocity only) can be switched on and off with no change whatsoever in the magnitude of the fields, but varying only their orientation in space. Therefore, the physical device being modeled in these cases can be adjusted to be either transparent or opaque by merely rotating the apparatus mechanically.

\section{Final remarks}\label{conclusion}
We studied some models for nonlinear electromagnetic material media which, in the regime of geometrical optics, do present phase transitions between optical transparency and opacity, and vice versa. These models were chosen to be mathematically simple and were mostly motivated by the present-day technological capability of designing and producing metamaterials whose properties can be set to meet pre-established requirements
\cite{Soukoulis2010}.

We do not present a complete analysis of this phenomenon in nonlinear media. Instead, we just present the method by which such phenomenon can be investigated. In this way, the theoretical description we present here could be a valuable tool in searching for new optical devices and effects. Depending on the external applied electric and magnetic fields, which couple to the dielectric properties of the medium by means its optical susceptibilities, opacity and transparency may occur in each studied model for certain directions of propagation of a probe electromagnetic wave. The models can be adapted to different situations, in order to yield the effect around specific directions of the wave propagation.

Another possible application of the present method would be the production of models presenting low values for light velocity around certain directions of propagation. A simple example is the model for electromagnetically switchable media investigated in Sec. \ref{electromag}, with a specific example depicted in Fig.\ \ref{fig3}. Propagation with directions near $\pi/2$ present quite low velocity. Slow light phenomena is an interesting tool for investigating some aspects of classical and quantum optics, including analog models of general relativity. In fact, in the context of analog models, slow light systems have been used to simulate black holes in several optical systems. The main idea is to measure the occurrence of the tiny Hawking radiation \cite{hawking1974}, long ago predicted to occur in astrophysical black holes, but its small magnitude renders its direct observation unlikely to occur in the realm of astrophysics. Thus, measuring this curious effect in terrestrial laboratories \cite{steinhauer2016} by means of analog models seems to be an important starting point to test semiclassical gravity predictions and, perhaps even quantum gravity phenomena.

Finally, it should be noticed that in all models here examined, except for the one studied in Sec. \ref{electromag}, birefringence phenomena occur; i.e., the solutions exhibit two distinct phase velocities along the same direction of the wave vector. The distinct solutions are associated with different polarization vectors. In all cases where the effect occurs its magnitude is larger at the center of the transparency windows and falls to zero at the borders.

\section*{APPENDIX: the canonical forms of the $[C_{ij}]$ matrix}
The branch of material media we are interested in here corresponds to the ones which admit only real elements for the dielectric coefficients. Consequently, the auxiliary matrices $[C_{ij}]$ and $[A_{ij}]$ possess only real entries, yielding the existence of at least one real eigenvalue, for each of them. However, their Jordan canonical form can be non-diagonal if the other eigenvalues are complex. Recall that it is enough to study the eigenvalues of $[C_{ij}]$ in order to know the properties of the generalized Fresnel matrix in the case of electrically switchable media with an opacity-to-transparency transition. For simplicity, we shall analyze here $[C_{ij}]$ solely. The decomposition in terms of the eigenvalues can also be applied to the magnetically switchable media.

Let us denote $\lambda_A\in\mathbb{C}$ the eigenvalues of the dimensionless version of $[C_{ij}]$ and a basis $\{\hat v^{(A)}_{\,j}\}$ of real vectors, where $A$ labels the vectors and $j$ labels the components, with $A,j=1,2,3$. If we restrict our analysis to the real canonical forms of $[C_{ij}]$, we have to consider two cases separately: either (a) $\lambda_{1,2}\in\mathbb{C}$ and $\lambda_3\in\mathbb{R}$ or (b) $\lambda_A\in\mathbb{R}$.

In case (a), $\lambda_1=x+yi$ with $y>0$ and the canonical form is
\begin{equation}
\label{can-com-c}
[C_{ij}]\big|_a=\left(
\begin{array}{ccc}
x&-y&0\\[1ex]
y&x&0\\[1ex]
0&0&\lambda_3
\end{array}\right),
\end{equation}
and for case (b), we obtain
\begin{equation}
\label{can-real-c}
[C_{ij}]\big|_{b}=\left(
\begin{array}{ccc}
\lambda_1&0&0\\[1ex]
\delta_1&\lambda_2&0\\[1ex]
0&\delta_2&\lambda_3
\end{array}\right),
\end{equation}
where $\delta_{1,2}=0,1$, depending on the eigenvalues degeneracy and the number of associated eigenvectors.

We then decompose the unit wave vector $\hat{k}_i$ in terms of the basis $\{\hat v^{(A)}_{\,j}\}$ as
\begin{equation}
\hat{k}_i=k_A\hat v^{(A)}_{\,i},
\end{equation}
where the coefficients $k_A\in\mathbb{R}$ lie upon the unit sphere $(k_1)^2 + (k_2)^2 + (k_3)^2=1$. Therefore, the coefficients of the polynomial equation for the phase velocity (\ref{coef-vp1})--(\ref{coef-vp3}) are
\begin{equation}
\label{coef-com-vp}
\left\{\begin{array}{l}
a=(x^2+y^2)\lambda_3,\\[1ex]
b=-\frac{1}{\mu}\left[(x^2+y^2)(1-k_{3}^2)-2y^2 k_{2}^2 + x\lambda_3(1+k_{3}^2)\right],\\[1ex]
d=\frac{1}{\mu^2}\left[x(1-k_{3}^2) + \lambda_3 k_{3}^2\right],
\end{array}\right.
\end{equation}
for case (a) and
\begin{equation}
\label{coef-real-vp}
\left\{\begin{array}{l}
a=\lambda_1\lambda_2\lambda_3,\\[1ex]
b=-\frac{1}{\mu}[\lambda_1\lambda_2(k_1^2+k_2^2) + \lambda_1\lambda_3(k_1^2+k_3^2) \\[1ex]
+\lambda_2\lambda_3(k_2^2+k_3^2) + \lambda_1\delta_2\,k_2k_3+\lambda_3\delta_1\,k_1k_2-\delta_1\delta_2\,k_1k_3],\\[1ex]
d=\frac{1}{\mu^2}(k_1^2\lambda_1 + k_2^2\lambda_2 + k_3^2\lambda_3+\delta_1 k_1k_2+\delta_2 k_2 k_3),
\end{array}\right.
\end{equation}
for case (b).

Finally, we display the discriminants of each case. For case (a),
\begin{eqnarray}
\Delta\big|_a&=&(1 - k_3^2)^2 x^4 - 2 (1 - k_3^2)^2 \lambda_3 x^3 + \big[(1 - k_3^2) \lambda_3^2 \nonumber\\
&&+2(1 - 2k_2^2 - k_3^2)y^2\big] (1-k_3^2) x^2 + 2\big[2(k_3^2-1) \nonumber\\
&&+(1 - 2k_2^2 - k_3^2)(1 + k_3^2)\big]\lambda_3y^2x - 4k_3^2\lambda_3^2 y^2\nonumber\\
&&+ (1 - 2k_2^2 - k_3^2)^2y^4.
\end{eqnarray}
For case (b), with $\lambda_1\neq\lambda_2\neq\lambda_3$, we obtain
\begin{eqnarray}
\Delta\big|_{b_{1}}&=&\left[\lambda_2(\lambda_1-\lambda_3) k_3^2+\lambda_1(\lambda_2-\lambda_3)\right]^2\nonumber\\
&&+\left[2 \lambda_2(\lambda_1-\lambda_3) k_3^2 + (\lambda_1-\lambda_2) \lambda_3 k_2^2\right.\nonumber\\
&&- 2 \lambda_1(\lambda_1-\lambda_2)\big] \lambda_3(\lambda_1-\lambda_2) k_2^2.
\end{eqnarray}
The most degenerate situation for the case (b) is $\lambda_1=\lambda_2=\lambda_3$ with only one eigenvector. Then, the discriminant is
\begin{equation}
\begin{array}{lcl}
\Delta\big|_{b_{2}}&=&[k_2(k_1 + k_3)\lambda_3 - k_1k_3]^2 - 4k_1k_3\lambda_3^2.
\end{array}
\end{equation}
Therefore, the negativity of each of these quantities guarantees the opacity of the medium.

\section*{ACKNOWLEDGEMENTS}
This work was partially supported by the Brazilian research agencies CAPES (Grant 1614969/PNPD) and CNPq (Grant 302248/2015-3).


\begin{thebibliography}{88}
%
\bibitem{meta}
J. B. Pendry, D. Schurig, D. R. Smith, Controlling electromagnetic fields, {Science} {\bf 312}, 1780 (2006).
%
\bibitem{harris1990}
S. E. Harris, J. E. Field, A. Imamoglu, Nonlinear optical processes using electromagnetically induced transparency, {Phys.\ Rev.\ Lett.} {\bf 64}, 1107 (1990).
%
\bibitem{fleis}
M. Fleischhauer, A. Imamoglu, J. P. Marangos, Electromagnetically induced transparency: Optics in coherent media, Rev.\ Mod.\ Phys. {\bf 77}, 633 (2005).
%
\bibitem{veselago1968}
V. G. Veselago, Electrodynamics of substances with simultaneously negative values of $\epsilon$ and $\mu$, {Sov. Phys. Usp.} {\bf 10}, 509 (1968).
%
\bibitem{smith2000}
D. R. Smith, W. J. Padilla, D. C. Vier, S. C. Nemat-Nasser, and S. Schultz, Composite medium with simultaneously negative permeability and permittivity, {Phys. Rev. Lett.} {\bf 84}, 4184 (2000).
%
\bibitem{jalas}
D. Jalas, A. Petrov, M. Eich, W. Freude, S. Fan, Z. Yu, R. Baets, M. Popovic', A. Melloni, J. D. Joannopoulos, M. Vanwolleghem, C. R. Doerr, and H. Renner, What is---and what is not---an optical isolator, Nat. Photon. {\bf 7}, 579 (2013).
%
\bibitem{pereira2014}
V. A. De Lorenci and J. P. Pereira, One-way propagation of light in Born-Infeld-like metamaterials, Phys.\ Rev.\ A {\bf 89}, 043822 (2014).
%
\bibitem{jonas2016}
J. P. Pereira, I. I. Smolyaninov, and V. N. Smolyaninova, Magnetic liquids under high electric fields as broadband optical diodes, Phys. Rev. A {\bf 94}, 043852 (2016).
%
\bibitem{tsakmakidis2007}
K. L. Tsakmakidis, A. D. Boardman, and O. Hess, ``Trapped rainbow'' storage of light in metamaterials, Nature (London) {\bf 450}, 397 (2007).
%
\bibitem{smolyaninov2010}
V. N. Smolyaninova, I. I. Smolyaninov, A. V. Kildishev, and V. M. Shalaev, Experimental observation of the trapped rainbow, Appl. Phys. Lett. {\bf 96} 211121 (2010).
%
\bibitem{Boller}
K. -J. Boller, A. Imamoglu, and S. E. Harris, Observation of electromagnetically induced transparency, Phys.\ Rev.\ Lett. {\bf 66}, 2593 (1991).
%
\bibitem{Hu}
Y. Hu, W. Liu, Y. Sun, X. Shi, J. Jiang, Y. Yang, S. Zhu, J. Evers and H. Chen, Electromagnetically-induced-transparency--like phenomenon with resonant meta-atoms in a cavity, Phys.\ Rev.\ A {\bf 92}, 053824 (2015).
%
\bibitem{tassin2012}
P. Tassin, L. Zhang, R. Zhao, A. Jain, T. Koschny, and C. M. Soukoulis, Electromagnetically induced transparency and absorption in metamaterials: The radiating two-oscillator model and its experimental confirmation, Phys.\ Rev.\ Lett. {\bf 109}, 187401 (2012).
%
\bibitem{Zhang}
S. Zhang, D. A. Genov, Y. Wang, M. Liu, and X. Zhang, Plasmon-induced transparency in metamaterials, Phys.\ Rev.\ Lett. {\bf 101}, 047401 (2008).
%
\bibitem{tassin2009}
P. Tassin, L. Zhang, T. Koschny, E. N. Economou, and C. M. Soukoulis, Low-loss metamaterials based on classical electromagnetically induced transparency, Phys.\ Rev.\ Lett. {\bf 102}, 053901 (2009).
%
\bibitem{kurter2011}
C. Kurter, P. Tassin, L. Zhang, Th. Koschny, A. P. Zhuravel, A. V. Ustinov, S. M. Anlage, and C. M. Soukoulis, Classical analogue of electromagnetically induced transparency with a metal-superconductor hybrid metamaterial, Phys.\ Rev.\ Lett. {\bf 107}, 043901 (2011).
%
\bibitem{stewart2013}
S. D. Jenkins and J. Ruostekoski, Metamaterial Transparency induced by cooperative electromagnetic interactions,
{ Phys. Rev. Lett.} {\bf 111}, 147401 (2013).
%
\bibitem{sm-glass}
R. Baetens, B. P. Jelle and A. Gustavsen, Properties, requirements and possibilities of smart windows for dynamic daylight and solar energy control in buildings: A state-of-the-art review, Solar Energy Mater.\ Solar Cells {\bf 94}, 87 (2010).
%
\bibitem{he}
X. He, X. Yang, S. Li, S. Shi, F. Wu, J. Jiang, Electrically active manipulation of electromagnetic induced transparency in hybrid terahertz metamaterial, Opt.\ Mat.\ Express {\bf 6}, 3075 (2016).
%
\bibitem{yasir}
K. A. Yasir and W.-M. Liu, Controlled Electromagnetically induced transparency and Fano resonances in hybrid BEC-optomechanics, Sci.\ Rep.\ {\bf 6}, 22651 (2016).
%
\bibitem{hadamard}
J. Hadamard, \textit{Le\c cons sur la propagation des ondes et les \'equations de hydrodynamique} (Hermann, Paris, 1903); V. D. Zakharov, \textit{Gravitational waves in Einstein's theory} (John Wiley \& Sons, New York, 1973).
%
\bibitem{dante}
D. D. Pereira and R. Klippert, Local nonlinear electrodynamics, Phys.\ Lett.\ A {\bf 374}, 4175 (2010).
%
\bibitem{local}
V. A. De Lorenci and R. Klippert, Electromagnetic light rays in local dielectrics, Phys.\ Lett.\ A {\bf 357}, 61 (2006).
%
\bibitem{goulart2008}
V. A. De Lorenci and G. P. Goulart, Magnetoelectric birefringence revisited, Phys.\ Rev.\ D {\bf 78}, 045015 (2008).
%
\bibitem{edu2012}
M. Novello and E. Bittencourt, Gordon metric revisited, Phys.\ Rev.\ D {\bf 86}, 124024 (2012).
%
\bibitem{jonas}
E. Bittencourt, J. P. Pereira, I. I. Smolyaninov, V. N. Smolyaninova, The flexibility of optical metrics, Class.\ Quantum Grav. {\bf 33}, 165008 (2016).
%
\bibitem{teodoro}
V. A. De Lorenci, R. Klippert and D. H. Teodoro, Birefringence in nonlinear anisotropic dielectric media, Phys.\ Rev.\ D {\bf 70}, 124035 (2004).
%
\bibitem{delorenci2002}
V. A. De Lorenci and R. Klippert, Analogue gravity from electrodynamics in nonlinear media, Phys.\ Rev.\ D {\bf 65}, 064027 (2002);
%
\bibitem{delorenci2002b}
V. A. De Lorenci, Effective geometry for light traveling in material media, Phys.\ Rev.\ E {\bf 65}, 026612 (2002).
%
\bibitem{barcelo2005}
C. Barcelo, S. Liberati, M. Visser, Analogue gravity, Living Rev.\ Relativity {\bf 8}, 12 (2005).
%
\bibitem{hawking1974}
S. W. Hawking, Black hole explosions, Nature (London) {\bf 248}, 30 (1974).
%
\bibitem{steinhauer2016}
J. Steinhauer, Observation of quantum Hawking radiation and its entanglement in an analog black hole, Nature Physics {\bf 12}, 959 (2016).
%
\bibitem{Soukoulis2010}
C. M. Soukoulis and M. Wegener, Optical metamaterials--more bulky and less lossy, Science {\bf 330}, 1633 (2010).

\end{thebibliography}
\end{document}